# Spatially Resolving Electron Spin Resonance of π-Radical in Single-molecule Magnet

*Ryo Kawaguchi\* [a], Katsushi Hashimoto [b,c], Toshiyuki Kakudate [d], Keiichi Katoh [e], Masahiro Yamashita [f,g], and Tadahiro Komeda\* [a,c].*

[a] Institute of Multidisciplinary Research for Advanced Materials (IMRAM, Tagen), Tohoku University, Sendai 980-0877, Japan

[b] Department of Physics, Graduate School of Sciences, Tohoku University, Sendai 980-8578, Japan

[c] Center for Science and Innovation in Spintronics (Core Research Cluster), Tohoku University, Sendai 980-8577, Japan

[d] Department of Industrial Systems Engineering National Institute of Technology (KOSEN), Hachinohe College, 16-1 Uwanotai, Tamonoki, Hachinohe, 039-1192, Japan

[e] Department of Chemistry, Graduate School of Science, Josai University, Sakado, Saitama 350-0295, Japan

[f] Department of Chemistry, Graduate School of Science, Tohoku University, Sendai 980-8578, Japan

[g] School of Materials Science and Engineering, Nankai University, Tianjin 300350, China

The spintronic properties of magnetic molecules have attracted significant scientific attention. Special emphasis has been placed on the qubit for quantum information processing. The single molecule magnet, bis(phthalocyaninato (Pc)) Tb(III) (TbPc$_2$), is one of the best examined cases in which the delocalized π-radical electron spin of the Pc ligand plays the key role in reading and intermediating the localized Tb spin qubits. We utilized the electron spin resonance (ESR) technique implemented on scanning tunneling microscope (STM) and use it to measure local ESR of single TbPc$_2$ molecule decoupled from the Cu(100) substrate by 2 monolayers NaCl film to identify the π-radical spin. We detected the ESR signal at the ligand positions at the resonance condition expected for the $S$ = 1/2 spin. The results reveal that the π-radical electron is delocalized within the ligands and exhibits intramolecular coupling susceptible to the chemical environment.

The spintronics technique exploiting spin-charge conversion is one of the important techniques used to develop electronic devices based on spin-quantum bits (qubits) to achieve quantum information processing (QIP)[1,2]. The ideas of quantum spintronics were proposed at early stages[3,4] and has been extensively used to electrically read-out the quantum information of single spin qubits implemented in the forms of quantum dots[5], dopants[6] in semiconductors, and more recently, in a form of a new class of systems referred to as single-molecule magnets (SMMs). The SMM has great potential for the realization of single-molecule qubits because it preserves the magnetization information for a specific time below the blocking temperature[7]. Specifically, one of the lanthanide (Ln)-based, double-decker Pc SMMs of the TbPc$_2$ molecule attached in nano-gap electrodes successfully demonstrated the quantum manipulation and read-out of the hyperfine coupled electronic ground state $J = \pm 6$ and nuclear spin $I = 3/2$ of the Tb ion through electric transport[8–10]. The mechanism of this spin-charge conversion was claimed to be based on the key role of the π-radical as the interface between the Tb spin and the conduction electron, manifested by the weak magnetic coupling of the delocalized π-radical electron spin with the localized magnetic moment of Tb $4f^8$ electrons. These magnetic coupling and delocalization are crucial for designing suitable interfaces for not only spin-charge conversion, but also one-dimensional (1D)[11–13] or two-dimensional (2D)[14,15] spin network to achieve QIP; thus, demanding a spatially resolved magnetic characterization of the delocalized π-radical spin within the individual SMM.

Scanning tunneling microscopy (STM) and spectroscopy (STS) revealed the existence of the spin of the TbP$_2$ molecule adsorbed on the metal surface based on the observation of the Kondo resonance[16]. The Kondo resonance appears owing to the spin screening by the conduction electron that forms the Kondo cloud. However, the information obtained by the Kondo resonance is limited compared with that obtained by a more direct method, such as electron spin resonance (ESR). For example, the Kondo resonance cannot provide direct evidence to attribute its origin to the π-radical spin of $S = 1/2$. Recently, the ESR technique was used in combination with STM (ESR-STM) to identify a magnetic moment in an individual atom at the high-energy resolution (nano-eV range)[17,18], which is not easily achieved in STS-based Kondo measurements[19–21].

In addition, several STS studies found that the interaction of the π-radical with the substrate alters the intrinsic spin characteristics of the SMM. For instance, the ligand spin is quenched when the TbPc$_2$ molecule is adsorbed on the Co[22] or Ag[20] surfaces. This effect is suppressed by an ultrathin dielectric film of MgO used as an insulating film. This has been demonstrated by ESR-STM measurements that have resolved the intrinsic spin of atoms[17,23–25]. However, a recent ESR-STM study showed that the spin character of a magnetic molecule, such as iron phthalocyaninato (FePc) molecules, is modified by charge transfer from the metal substrate even when protected by the MgO film[21,26]. A similar case was also observed in π-radical-based molecules ($C_{22}H_{14}$) deposited on a MgO/Ag substrate[27]. It is necessary to examine other types of dielectric films to highlight the intrinsic spin characteristics of the π-radical electron at the single molecular level.

In this study, we performed ESR-STM measurements on a single $TbPc_2$ molecule adsorbed on dielectric 2 monolayer (ML) NaCl films which covered a Cu(100) surface. Even though the MgO film is often employed in ESR-STM studies[17,23–25,28], we deliberately used 2ML NaCl that has a different and a more versatile nature as an insulating film. The radio frequency (RF) signal was superimposed on the DC tunneling bias voltage ($V_{DC}$), and the magnetic field was generated by a superconducting magnet. The STM tip was decorated with a few Fe atoms to be spin-polarized. The ESR spectrum described as the RF-induced tunneling current versus RF frequency showed a sharp ESR signal only upon measurements at the Pc ligand positions within the single $TbPc_2$ molecule. The peak frequency exhibited a linear dependence on the magnetic field and yielded a $g$-factor equal to 1.84 ± 0.12. The resonance behavior demonstrates that we spatially and energetically resolved the $S = 1/2$ π-radical electron spin inherent to the Pc ligand of the single $TbPc_2$ molecule. Furthermore, an intriguing difference appeared from the conventional ESR results, whose mechanism will be discussed in the following section.

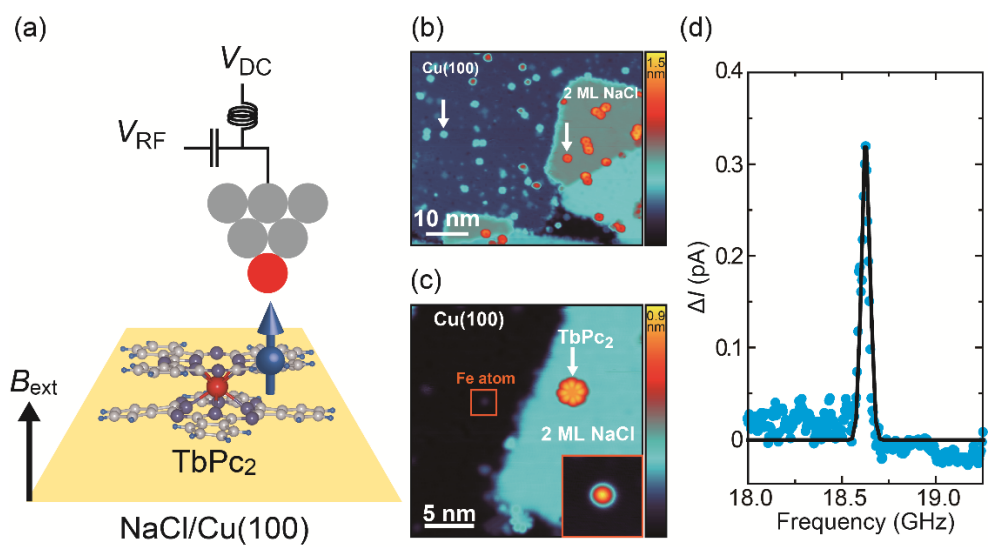

**Figure 1.** Electron spin resonance-scanning tunneling microscopy (ESR-STM) measurements of single TbPc$_2$ molecules adsorbed on a NaCl/Cu(100) surface. (a) Schematics of the experimental setup. Single TbPc$_2$ molecule adsorbed on 2 ML NaCl/Cu(100) surface is studied using ESR-STM with an out-of-plane external magnetic field ($B_{ext}$). Radiofrequency (RF) voltage ($V_{RF}$) is applied to the tunneling junction with a DC bias voltage ($V_{DC}$) based on the use of a bias Tee. A few Fe atoms (represented by red-filled circles) are picked up by the tip apex from the Cu surface to prepare the spin-polarized magnetic tip. (b) Large-scale constant current STM image of TbPc$_2$ molecules adsorbed on the 2 ML NaCl/Cu(100) and Cu(100) surfaces. Scanning conditions: $V_{DC}$ = −1.5 V, $I$ = 24 pA. The white arrows indicate representative isolated TbPc$_2$ molecules. (c) Small-scale constant current STM image of a representative isolated single TbPc$_2$ molecule (marked by a white arrow) adsorbed on the 2 ML NaCl/Cu(100) surface. The eight bright lobes correspond to the upper Pc ligand and darker center of the Tb ion. The orange square marks a Fe atom. Scanning conditions: $V_{DC}$ = −1.3 V and $I$ = 10 pA. (d) ESR spectrum measured at a lobe position of the TbPc$_2$ molecule on the 2 ML NaCl/Cu(100) surface with an external out-of-plane magnetic field

of 650 mT. ESR-STM conditions: $V_{DC}$ = −100 mV, $I$ = 10 pA, $V_{RF}$ = 10 mV. The ESR spectrum is fitted with a Lorentzian function (black curve).

The measurements were conducted with the use of a low-temperature $^3$He STM system (Unisoku UHV USM1300) equipped with a superconductor magnet and high-frequency semi-rigid cable which connected the tip bias line (more details of the measurement setup are provided in the Supporting Information (SI), Section 1). Before the ESR-STM measurements, we prepared the spin-polarized STM tip by picking up a few Fe atoms from the Cu surface with the STM tip and then confirmed the spin polarization determined by the asymmetric *dI/dV* spectrum of the TbPc$_2$ molecule (see SI Section 2). Figure 1(a) illustrates the experimental setup of ESR-STM to investigate the spin resonance of the single-molecule of TbPc$_2$. The RF signal voltage ($V_{RF}$) is superimposed to the DC tunneling bias voltage ($V_{DC}$) by using the bias Tee, followed by its use in the tunneling junction via the semi-rigid and flexible coaxial cables (see SI Section 1). The frequency-dependent transmission of the RF line, which was determined based on the use of the surface state of the Ag(111) surface[29], was compensated by tuning the RF source power (more details of the compensation technique are provided in SI, Section 3) to apply the constant amplitude of the RF voltage ($V_{RF}$) at the tunneling junction. The tunneling current enhancement at the resonance condition was detected using the lock-in technique in which the RF signal was chopped at a frequency of 431 Hz, and the synchronized tunneling current variation was detected by the lock-in amplifier (see SI Section 3). To decouple the TbPc$_2$ molecules from a metal substrate electronically, we selected the dielectric NaCl thin film[30–32]. The NaCl film was deposited by thermal evaporation from a Knudsen cell (K-cell) onto the Cu(100) surface at room temperature, which metal substrate was precleaned following several cycles of Ar+ ion sputtering and electron bombardment annealing at 670 K. Subsequently, TbPc$_2$ molecules were deposited onto the substrate at 50 K from another K-cell. Additionally, Fe atoms were deposited onto the substrate at 30 K by electron-bombardment evaporation to prepare the spin-polarized tip (See SI Section 2). All ESR measurements were completed by minimizing the tunneling current ($I$ = 10 pA) and with the application of $V_{DC}$ = −100 mV to the tip.

Figure 1(b) shows the STM topographic image of the surface prepared by depositing 2 monolayers (ML) of NaCl film onto the Cu(100) surface followed by the deposition of a submonolayer amount of TbPc$_2$. On the right side of the STM image, 2 ML NaCl film partially covering Cu(100) was visible, while a bare Cu(100) surface was exposed in other parts. Among the coinage metals, the adsorption energy of TbPc$_2$ was higher on the Cu surface than on Ag and Au[33]. Accordingly, an isolated TbPc$_2$ molecule can be observed easily on the Cu(100) surface (marked by a white arrow, see the left side). The NaCl film provides a smaller adsorption energy when used in conjunction with a TbPc$_2$ molecule, and the molecule hops on the film and tends to form a dimer or larger cluster. As a result, we can barely see an isolated molecule on the NaCl film as indicated by the white arrow; however, the number of the molecules is equal to 1/3 of that on the Cu surface, as estimated by the statistical analysis over a wider areas (see SI Section 4). In the magnified STM image of Figure 1(c), the TbPc$_2$ molecule shows a characteristic eight lobe-like shape which appearing on the side of the benzene ring of the Pc ligand. In addition, the Fe atom (indicated by the orange-colored square) deposited to prepare the spin-polarized tip are visible in the optimized color-scale STM topographic image (inset).

We performed ESR-STM measurements on the isolated single TbPc$_2$ molecule adsorbed on the 2 ML NaCl/Cu(100) surface based on the following measurement conditions: $V_{RF}$ = 10 mV, $I$ = 10 pA, and $V_{DC}$ = −100 mV; these were used in all ESR-STM measurements. Figure 1(d) shows the tunneling current variation ($\Delta I$) obtained on one of the lobes of the TbPc$_2$ molecule by sweeping the RF frequency from 18.0 GHz to 19.3 GHz under the out-of-plan magnetic field $B_{ext}$ = 650 mT, in which a sharp peak around 18.6 GHz exhibits the tunneling current enhancement $\Delta I$ = 0.3 pA. By fitting the spectrum with a Lorentzian function (black curve), we determined the linewidth of the peak of 55 MHz and quantified the energy resolution of approximately 100 nano-eV at this ESR-STM condition. The observed peak of the tunneling current appeared only for the TbPc$_2$ molecule adsorbed on the 2 ML NaCl(100)/Cu(100) surface but not on the Cu(100) surface. This molecule-derived peak was attributed to the ESR driven in the single TbPc$_2$ molecule according to the following process. The $V_{RF}$ modulated the magnetic-field gradient generated by the spin-polarized tip. At the $V_{RF}$ frequency corresponding to the Zeeman energy lifted by the external magnetic field, the effective oscillating magnetic field imposed by $V_{RF}$ drives the ESR transition between the Zeeman split states. This resulted in changes in the population of the splitting states, and then altered the tunneling magnetoresistance, which can be detected as a tunneling current variation $\Delta I$ with the use of a spin-polarized tip[17].

To examine the origin of the resonance peak, we investigated the position-dependent variation of the ESR-STM spectra within the molecule. Figure 2(a) shows the ESR-STM spectra obtained in the two different lobes and at the center of the TbPc$_2$ molecule, as illustrated by the circles superimposed on the topographic image of the molecule (Figure 2 (b)) with colors that match those of the spectra. The apparent peaks are observed at the same frequency at two different lobe positions, whereas the peak was absent at the center position. This is similar to what we observed previously for the spin-dependent Kondo distribution in the TbPc$_2$ molecule adsorbed on Au(111), in which the sharp Kondo peak was observed at the lobe position, while it disappeared at the center position[16]. These results suggest that our ESR-STM probed the $S = 1/2$ $\pi$-radical electron spin that was distributed on the Pc ligands near the lobe positions, as schematically drawn by the blue ring in Figure 2(c).

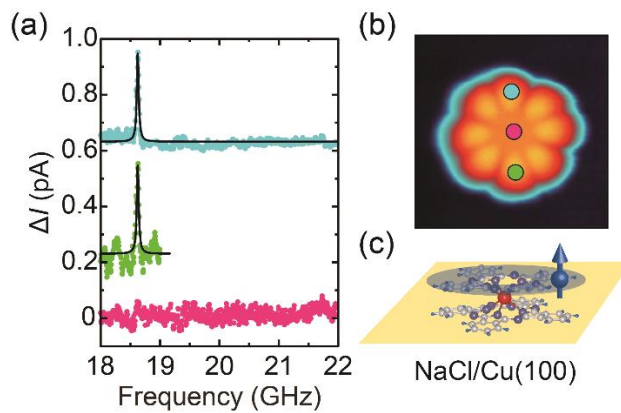

**Figure 2.** Spatial distribution of ESR signal within a single TbPc$_2$ molecule. (a) ESR spectra measured at the positions marked with the corresponding color in (b) at $B_{ext}$ = 650 mT. All spectra were acquired at the same conditions: $V_{DC}$ = −100 mV, $I$ = 10 pA, and $V_{RF}$ = 10 mV. Each spectrum is shifted vertically for improved visualization. Black curves superimposed on the two spectra are Lorentzian fits. (b) STM image of single TbPc$_2$ molecule adsorbed on the 2 ML NaCl/Cu(100) surface. Scanning conditions: $V_{DC}$ = −100 mV and $I$ = 10 pA. (c) Schematics of the π-radical electron spin distribution (blue ring) within the TbPc$_2$ molecule adsorbed on the 2 ML NaCl/Cu(100) surface.

To clarify that the detected peak is the resonance of the π-radical electron spin, we obtained spectra at the same lobe position at different external magnetic fields ($B_{ext}$ = 650, 750, and 800 mT), and showed them in Figure 3(a). As shown, the ESR peak shifts to a higher frequency with an increasing magnetic field. The determined ESR frequencies are plotted as a function of $B_{ext}$ in Figure 3(b), thus indicating the linear Zeeman shift described by $hf = g\mu_B S B_{ext}$. Here, h is the Planck's constant, $f$ is the resonance frequency, and $\mu_B$ is Bohr magnetron. The g-factor was obtained from the linear fitting of the plot in Figure 3(b) and is equal to 1.84 ± 0.12. In addition, we determined the zero-field intercept $f_0$ to be 1.8 ± 1.0 GHz using the Figure 3(b).

We interpreted the results by considering the frequency–magnetic-field relation described for the resonance condition by the following Hamiltonian[34,35].

$H = \mu_B B_{ext}(g_{Ln}J^z_{Ln} + g_{rad}S_{rad}) - J_{ex}(J^z_{Ln} \cdot S_{rad})$ (1)

Here, $g_{Ln}$ and $g_{rad}$ represents the g-factor for the lanthanide and radical, respectively. The first term is the Zeeman energy. In the case of Ln=Tb, $J^z_{Tb}$ and $S_{rad}$ represent z component of the total angular momentum of Tb ion and the spin of the radical, which attain a value of 6 and 1/2, respectively. The expected values for $g_{rad}$ and $g_{Tb}$ are 2 and 3/2, respectively[36]; in particular, the $g_{rad}$ was experimentally confirmed by the conventional ESR measurement for an ensemble of TbPc$_2$ molecules, such as single crystal or powder sample[37]. The obtained value of g-factor (= 1.84) is closer to g = 2 than g = 3/2, which lends support to the detection of a π-radical feature in our experiment. On the other hand, our result indicates finite deviation for the isolated TbPc$_2$ on the NaCl/Cu substrate from g = 2 obtained for the ensemble TbPc$_2$ molecules. However, we can see the similar deviation in the g-factor for the lanthanide-radical complexes. For example, the [Ln(hfac)$_3$(TMIO)$_2$] (Ln = Tb, Dy; TMIO = 1,1,3,3-tetramethylisoindolin-2-oxyl; hfac = 1,1,1,5,5,5-hexafluoropefntane-2,4-dionate) molecule shows g = 1.82 for Ln=Tb[35], and g = 1.92[38] for Tb$_2$pyNO (*tert*-butyl 2-pyridyl nitroxide). Moreover, a substantially large $g = 2.57$ was observed for the M$^{III}$(Pc)(TNP) (M=Tb and Dy; Pc = phthalocyaninate, TNP = tetranaphthylporphyrinate)[34], even though the electronic structure of porphyrin and its derivatives are similar to that of Pc as a ligand.

The deviation of *g*-factor from 2 can be explained by mixing of the $g_{Tb}$ component in the $g_{rad}$ [34,35]. The significant deviation observed in the case of Tb(Pc)(TNP), which is absent for the TbPc$_2$ molecule possessing the higher-symmetric structure, is related to the molecule symmetry; a larger mixing of the radical and the lanthanide state is expected for a lower molecular symmetry[35]. We therefore attribute the *g*-factor deviation obtained in this study to the mixture of the Tb field. The mixing can be realized by potential distribution, which is different from the single crystal, as discussed later.

The second term in Eq. (1) corresponds to the zero-field intercept, $f_0$. The $f_0$ determined from in this experiment is significantly smaller than that observed for the single crystal. Note that the $f_0$ differences cannot be attributed to experimental artifacts. In the ESR-STM setup, the stray magnetic field from the spin-polarized tip, the tip field ($B_{tip}$), can be superimposed on $B_{ext}$. However, from the tunneling current of 10 pA used in our measurements, $B_{tip}$ can only equal a few tens of mT[23–25]. Additionally, the magnetic hysteresis of the superconductor magnet can yield an error approximately equal to ±10 mT. These are too small to account for large deviations of zero-field intercept from the ensemble ESR. We argue that our observed small intercept of $f_0 = 1.8 \pm 1.0$ GHz is due to the exchange interaction in the isolated single TbPc$_2$ molecule being significantly weaker than that of the ensemble case.

The possible cause for the weak exchange interaction is the chemical environment of the individual molecule. Specifically, for the single molecule adsorbed on a substrate, the intramolecular electronic orbital is affected by the coupling with the substrate, while being free from the intermolecular interactions relevant to the molecule in the bulk crystal. As a result, the magnetic properties are no longer identical to that in the bulk crystal. In the current experiment, the isolated single molecule is electrically decoupled from the metal substrate by the insulating film of NaCl which suppresses the influences of the metal substrate, such as charge transfer or hybridization. Thus, the NaCl film preserves the inherent electric properties of the free-standing molecule[30].

Conversely, owing to the dielectric nature of the NaCl film, the long-range polarization effect should deform the electronic structure of the individual TbPc$_2$ molecule in a manner different from that of the TbPc$_2$ bulk crystal. A theoretical study showed that upon adsorption on NaCl/Ge, the energy separation of the carbon monoxide molecule HOMO-LUMO gap changes depending on the thickness of the NaCl layer and is accounted for by the long-range polarization effect originating from the NaCl/Ge interface[39]. This gap behavior has also been reported in STS experiments on pentacene molecules adsorbed on a NaCl/Cu surface[40]. It is also argued that the exchange interaction energy is sensitive to the intramolecular charge distribution[41]. From these previous works, we consider the small exchange interaction between the radical and the Tb (and the small $f_0$) is derived from the long-range polarization effect induced by the NaCl/Cu interface and/or the absence of intermolecular interaction in an isolated state on the NaCl.

The polarization effect induces asymmetric potential for the upper and lower Pc ligands in our experiment, which can cause a strong deviation of the *g*-factor from 2. The differences between the magnetic property of the single crystal and the film on an insulating layer are observed for the TbPc$_2$/MgO case in the magnetic remanence measurements, which can be explained by the potential difference between the two cases[42]. The effect of the MgO film on the *g*-factor deviation is also discussed in the ESR-STM study of the FePc molecule (in the SI of the Ref. [21,26]).

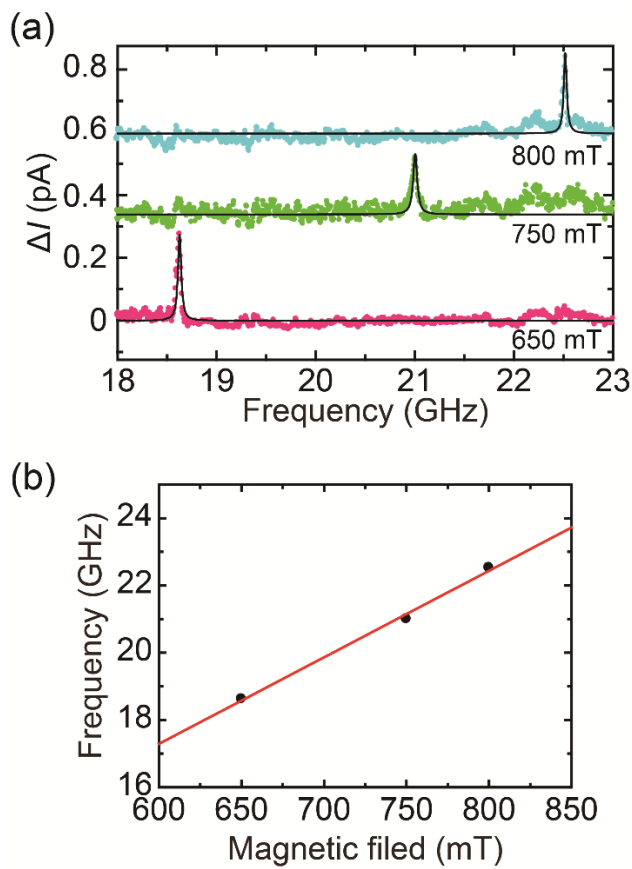

**Figure 3.** External magnetic-field dependence of the ESR spectra on the ligand of TbPc$_2$ molecule. (a) ESR spectra obtained at the same ligand position marked by one of the dots in Figure 2(b) and at the same ESR-STM conditions. Each spectrum is vertically shifted for improved visualization. The solid black curves are obtained from Lorentzian fits. (b) Magnetic field dependence of the resonance frequencies determined in (a).

Finally, we discuss the role of NaCl film as an insulating layer that decouples the electronic interaction from the metal substrate in the EPR-STM experiment. MgO film has been extensively employed[17,21,23,28,43,44] for this purpose, but we use NaCl film for the optical-STM experiments of single molecules[45,46]. One part of the reason for using the NaCl film is the less complicated film formation process as compared to the MgO film, while another part is the intriguing characteristics of the NaCl film. Even though MgO film is a good insulator, charge transfer between the adsorbed molecules and the substrate has been reported; the FePc molecule shows $S = 1/2$ on MgO film[21,26], different from the $S = 1$[47,48], as a result of charge transfer from the metal substrate. A similar case is also reported in other molecules[27]. We investigated such an effect with the NaCl film and found that at least the π-radical electron of TbPc$_2$ is preserved on the NaCl film, although it vanishes on other metal surfaces.

A DFT calculation showed the difference between the two films, where the Co atom's binding energy ($E_b$) is calculated on the two films[49]. The results reveal that $E_b$ is approximately two times higher on the MgO than on the NaCl surface. The intrinsic difference between the two insulating films can be traced back to the Madelung potential[49]. The MgO film consisting of divalent ion crystals yields a much higher value than monovalent NaCl. Thus, the NaCl film employed in our experiment forms a "soft" insulating film between the molecule and the metal substrate, suppressing the charge transfer and hybridization between them.

In conclusion, we demonstrated ESR-STM measurements on the single TbPc$_2$ molecule decoupled from the metal substrate; this was achieved by utilizing the 2ML NaCl film. We resolved π-radical $S = 1/2$ electron delocalization at the Pc ligands that possessed a $g$-factor equal to 1.84 ± 0.12. The $g$-factor and intercept observed in the plot of the magnetic field versus resonance frequency deviate from those for the TbPc$_2$ bulk crystal. The discrepancy can be explained by the intramolecular interaction susceptible to the chemical environment, which is different in bulk crystal molecules and surface-adsorbed single molecule on the NaCl film.

Our ESR-STM measurements in conjunction with the decoupling method can be applied to detect[50] and manipulate the hyperfine-coupled Tb spins in the TbPc$_2$ molecule with the minimization of decoherence owing to the coupling to the metal substrate. This can be extended to a study on the magnetic inter- and intramolecular interaction mediated by the π-radical electron in the 1D[11–13] or 2D[14,15] network, wherein the localized Tb spins can be potentially used as coupled qubits to achieve molecule-based QIP.

**Supporting Information**. The following files are available free of charge.

S1. Details of experimental setup

S2. Confirmation of spin-polarized tip with *dI*/*dV* spectrum

S3. Acquisition and compensation of transfer function

S4. Statistical analysis of the number of isolated TbPc$_2$ molecules on the surface (PDF)

**Corresponding Author**

**Ryo Kawaguchi** − Institute of Multidisciplinary Research for Advanced Materials (IMRAM, Tagen), Tohoku University, 2-1-1, Katahira, Aoba-Ku, Sendai 980-0877, Japan; https://orcid.org/0000-0002-1211-3550; Email: ryo.kawaguchi.a1@tohoku.ac.jp

**Tadahiro Komeda** − Institute of Multidisciplinary Research for Advanced Materials (IMRAM, Tagen), Tohoku University, 2-1-1, Katahira, Aoba-Ku, Sendai 980-0877, Japan; Department of Chemistry, Graduate School of Science, Tohoku University, Aramaki-Aza-Aoba, Aoba-Ku, Sendai 980-8578, Japan; https://orcid.org/0000-0003-2342-8184; Email: tadahiro.komeda.a1@tohoku.ac.jp

**Author Contributions**

R.K. conducted the measurements and data analysis and wrote the manuscript with advice of K.H. and T.K. R.K. developed the ESR-STM system with the help of T.K., K.H. K.K. and M.Y. synthesized the TbPc$_2$ molecules. All authors discussed the results and contributed to the writing of the manuscript.

**Funding Sources**

This study was supported in part by the Grant-in-Aid for Scientific Research (S) (No.19H05621) (R.K., K.H., T.K.) and by the National Natural Science Foundation of China (NSFC, 21971124, 22150710513). Special thanks are devoted for the support from the 111 projects (B18030) from China. K.H., T.K., and R.K. thank the TIA Kakehashi program for their support.

**Notes**

The authors declare no competing financial interests.

R.K. thanks Andreas Heinrich, Soo-Hyon Phark, and Fabio Donati for fruitful discussions.

# Supporting Information

# Spatially Resolving Electron Spin Resonance of π-Radical in Single-molecule Magnet

*Ryo Kawaguchi* [a], *Katsushi Hashimoto* [b,c], *Toshiyuki Kakudate* [d], *Keiichi Katoh* [e], *Masahiro Yamashita* [f,g], *and Tadahiro Komeda* [a,c].

[a] Institute of Multidisciplinary Research for Advanced Materials (IMRAM, Tagen), Tohoku University, Sendai 980-0877, Japan

[b] Department of Physics, Graduate School of Sciences, Tohoku University, Sendai 980-8578, Japan

[c] Center for Science and Innovation in Spintronics (Core Research Cluster), Tohoku University, Sendai 980-8577, Japan

[d] Department of Industrial Systems Engineering National Institute of Technology (KOSEN), Hachinohe College, 16-1 Uwanotai, Tamonoki, Hachinohe, 039-1192, Japan

[e] Department of Chemistry, Graduate School of Science, Josai University, Sakado, Saitama 350-0295, Japan

[f] Department of Chemistry, Graduate School of Science, Tohoku University, Sendai 980-8578, Japan

[g] School of Materials Science and Engineering, Nankai University, Tianjin 300350, China

S1. Details of experimental setup
S2. Confirmation of spin-polarized tip with $dI/dV$ spectrum
S3. Acquisition and compensation of transfer function
S4. Statistical analysis of the number of isolated TbPc$_2$ molecules on the surface

## S1. Details of the experimental setup

All experiments were performed in an ultrahigh-vacuum condition and at the sample temperature of 0.4 K achieved by the $^3$He cooling system (Unisoku USM1300). The cooling system was also equipped with a superconductor magnet (Cryogenic Ltd.) which was able to generate the out-of-plane magnetic field up to 11 T. We introduced the RF signal through the high-frequency cables connected to the tip bias line (Figure S1). The SMP-type feedthrough (Solid Sealing Technology, FA33179-1FF) create the connection between the ambient and the internal vacuum chamber (IVC), which had a lower insertion-loss performance than that of the standard SMA-type. We employed the semi-rigid cable of SC-219/50-SCN-CN (COAX Co., Ltd.) from this connector to the 1 K pot (orange line), and from the 1 K pot to the ultrahigh vacuum (UHV) feedthrough (green line), SC-119/50-SCN-CN (COAX CO., LTD). In the UHV chamber, the connection to the STM tip (blue line) was established by using the flexible coaxial cable (Cooner Wire, CW2040-3650F). For the line of the tunneling current, we used a SMA-type connector (COSMOTEC Co., C34SMR1) and flexible coaxial cable (same as that used for the RF line, as shown by the blue line in Figure S1). The continuous sinusoidal signal from the RF generator (Keysight, N5193A) was applied to the tunneling junction ($V_{RF}$) through the bias Tee (SigaTek, SB15D2) and combined with the DC bias voltage ($V_{DC}$). The cable shown with the yellow line in the figure is flexible coaxial cable (Mini-Circuits, TMP40-6FT-KMKM+).

For the lock-in detection, we modulated the amplitude of the RF signal (hence the RF bias voltage ($V_{RF}$)) in a chopping scheme at a frequency of $f_{mod}$ = 431 Hz. The resulting $V_{RF}$-induced tunnel current was detected with a lock-in amplifier (NF, LI5650) after an amplification by using the current amplifier (Femto, DLPCA-200) with the bandwidth which covered the $f_{mod}$. The lock-in output ($V_{lock-in}$) corresponded to the root-mean-square value of the $V_{RF}$-induced tunnel current;

it is defined as the $V_{RF}$-induced variation in the tunnel current ($\Delta I$). The ESR signal was acquired with a relatively weak feedback loop with the spin-polarized tip prepared by picking up a few Fe atoms from the Cu surface, while STM images were obtained at the constant-current mode.

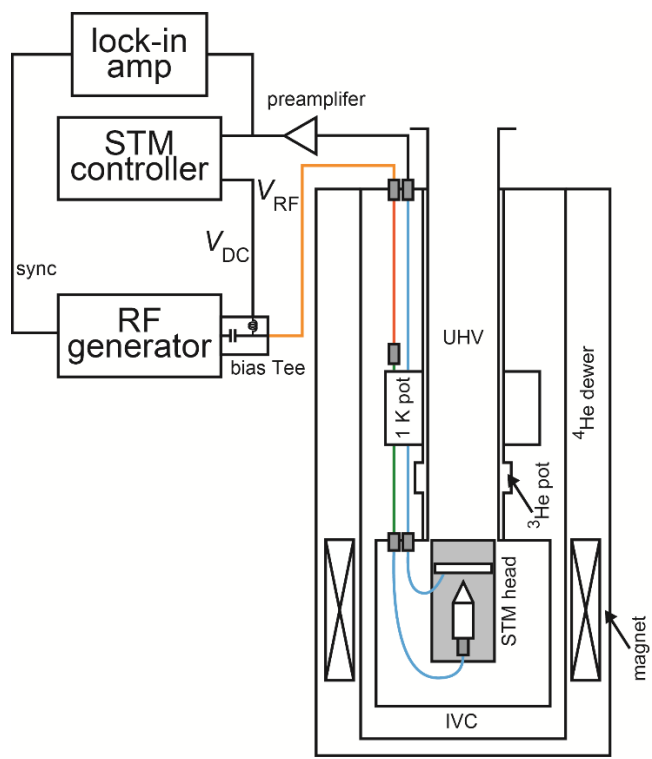

**Figure S1.** Schematic of radiofrequency (RF) cabling in the cryostat and outside of the measurement setup area used for electron spin resonance measurements.

## S2. Spin-polarized tip with *dI/dV* spectra

Figure S2(a) shows the *dI/dV* spectrum acquired with a nonmagnetic PtIr tip at the lobe position of the TbPc$_2$ molecule absorbed on the Cu(100) surface in the absence of the magnetic field. The sharp peak near the Fermi level corresponds to the Kondo peak. In the magnetic field ($B_{ext}$ = 500 mT), as shown by the black curve in Figure S2(b), this peak splits into two peaks marked by arrows, thus resulting in the symmetric spectrum concerning the bias polarity. The obtained splitting was caused by spin excitation inelastic tunneling spectroscopy (IETS) owing to the transition between the Zeeman split state $m = \pm 1/2$ of the π-radical electron of $S = 1/2$. Conversely, the *dI/dV* spectrum obtained with the spin-polarized tip, which was prepared by picking up a few Fe atoms from the Cu surface (red curve in Figure S2b), shows the asymmetric shape owing to the selection rule for spin excitation with tunneling electron[1,2]. By using this method, we confirmed the tip polarization before we conducted the ESR measurements.

The prepared spin-polarized tip is used in the ESR-STM measurements to read-out the population difference of the up and down spins, where difference between the on- and off-resonance conditions is given by the tunneling magnetoresistance (TMR). The tip also contributes to the driving of surface spin resonance.

(1) Read out of spin resonance state

The change in the population of the Zeeman splitting states due to the ESR was read out as the change of tunneling current ($\Delta I$) compared to the background tunneling current during the off-resonance condition between the spin polarized tip and π-radical electron spin on the surface. The tunneling conductance was determined by the relative spin alignment of spin polarized tip and π-radical electron due to the TMR described as[3]

$G = G_j(1 + \alpha \langle S_{tip} \rangle \cdot \langle S_\pi \rangle)$,

where α is a constant proportional to the tip polarization, $G_j$ is the spin averaged junction conductance, and $<S_{tip}>$ and $<S_\pi>$ are the spin value for tip and π-radical electron, respectively. The read out of spin state in the ESR-STM measurement can be understood as a modification of $S_\pi$ at the on-resonance condition created by RF signal. To read out the changing spin state, the RF bias voltage ($V_{RF}$) is applied to the tunneling junction with usual DC bias voltage ($V_{DC}$) in STM. The total applied bias voltage is

$V = V_{DC} + V_{RF}\cos(\omega t + \Phi)$,

where ω and Φ indicate the frequency and the phase of RF signal.

The tunneling current can be written as

$I(t) = GV = G_j(1+\alpha<S_{tip}>\cdot<S_\pi>)(V_{DC} + V_{RF}\cos(\omega t + \Phi))$.

Finally, the read out of ESR signal is determined by the difference in $I(t)$ between on- and off-resonance conditions of the π-radical electron spin.

(2) Driving the ESR of a target spin

The $V_{RF}$ modulated the magnetic-field gradient generated by the spin-polarized tip. At the $V_{RF}$ frequency corresponding to the Zeeman energy lifted by the external magnetic field, the effective oscillating magnetic field imposed by $V_{RF}$ drives the ESR transition between the Zeeman split states. The ESR-STM driving mechanism using the spin-polarized tip is understood by considering the spin interactions between the tip and the surface spin[4]. In ESR-STM measurement, the RF electric filed was applied from the spin-polarized tip to surface spin, whereas application of the oscillating magnetic field is generally required to drive the ESR. The mechanism of converting the RF electric field into the oscillating effective magnetic field is still under debate, and several experimental and theoretical investigations have been reported.

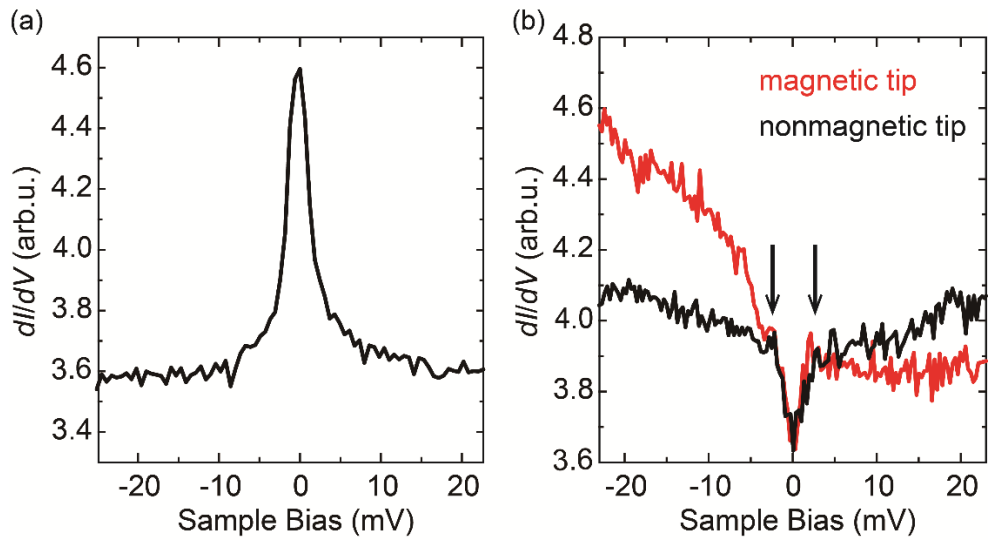

**Figure S2.** Preparation of spin-polarized scanning tunneling microscopy (STM) tip. (a) $dI/dV$ spectrum acquired at the lobe position of TbPc$_2$ molecule absorbed on the Cu(100) surface in the absence of a magnetic field. (b) $dI/dV$ spectra acquired with (red curve) and without (black curve) a magnetic tip in the presence of a magnetic field ($B_{ext}$ = 500 mT). All $dI/dV$ spectra were acquired at the same conditions ($V_{DC}$ = 25 mV, $I$ = 500 pA, $V_{mod}$ = 1 mV, $f_{mod}$ = 431 Hz).

## S3. Acquisition and compensation of transmission

The transmission of the entire RF signal line (Figure S1) has a specific frequency dependency owing to the characteristics of the RF components (such as cables, connectors, and feedthrough). As this frequency-dependent transmission makes the identification of the ESR peaks difficult, it is required to calibrate the transmission characteristics prior to ESR measurements. In frequency sweeping ESR-STM measurements with the constant RF source power ($P_{RF}$), frequency-dependent transmissions modify the amplitude of the RF voltage ($V_{RF}$) applied to the tunneling junction, and hence alter the ESR signal obtained as the tunnel current enhancement ($\Delta I$), thus leading to the artifact in the ESR spectrum. This artifact can be misinterpreted as an ESR signal, thus making it difficult to identify the actual ESR signal.

To compensate for the frequency-dependent influence of the transmission, we tuned $P_{RF}$ by the following procedure, as reported previously[5]. First, we quantified the $V_{RF}$ amplitude by rectifying the nonlinearity in the $I-V$ characteristic obtained from the Ag(111) surface state[6]. The Ag(111) surface was prepared with repeated Ar$^+$ ion sputtering and annealing at 773 K. As shown in the STM images (Figure S3(a)), the prepared surface shows the monolayer steps with hexagonal atomic structures (inset), which constitute evidence that an atomically clean surface was prepared. The $dI/dV$ spectrum obtained on the Ag(111) surface (blue spectrum in Figure S3(b)) shows the step-like structure and indicates the onset of the surface state. The application of the continuous wave RF signal ($P_{RF} = -5$ dBm, $f = 19$ GHz) broadens the step structure (black spectrum in Figure S3(b)). We fitted both spectra with an arcsine function and determined the $V_{RF}$ amplitude to be 25 mV at the specific RF condition. To evaluate the $V_{RF}$ variation as a function of $P_{RF}$, we first measured the $\Delta I$ based on the $V_{\text{lock-in}}$, as described in S1, while we swept $P_{RF}$ at a constant RF frequency ($f = 19$ GHz). The $V_{\text{lock-in}}$ was then scaled to $V_{RF}$ by using the relation between $V_{RF}$ and $V_{lock-in}$ obtained by the surface-state measurement described above. By using this method, we determined the $P_{RF}$ which corresponded to any required $V_{RF}$[5].

Next, we acquired the RF transmission on the Ag(111) surface by sweeping the frequency in the range of 18–25 GHz at a constant power setting equal to $P_{RF}$ = +5 dBm. The tunneling condition of $V_{DC}$ was adjusted to the Ag(111) surface state onset of −70 mV (see Figure S3(b)) during the application of an RF signal for rectification of nonlinearity of the I–V curve. Herein, we set the tunneling current to 500 pA, which was higher than that used in our ESR measurement (10 pA). The higher tunneling current increased the $V_{RF}$ and thus allowed us to obtain a more accurate relative intensity of the RF signal over the entire frequency range, especially in the frequency region in which the signal was strongly attenuated within the RF line. The RF transmission can be determined by dividing $V_{lock-in}$ by the source power of $P_{RF}$. Figure S3(c) shows the RF transmission measured on an Ag(111) surface in the frequency range of 18–25 GHz at a constant $P_{RF}$ = +5 dBm.

Finally, to compensate for the RF transmission, we tuned $P_{RF}$ (as plotted in Figure S3(d)) which was calculated from the obtained transmission function based on a method described previously[5]. The result (Figure S3(e)) obtained on the NaCl/Cu(100) surface demonstrates that we can produce a constant $V_{RF}$ = 5 mV over the frequency range of 18–25 GHz used in our ESR-STM measurements.

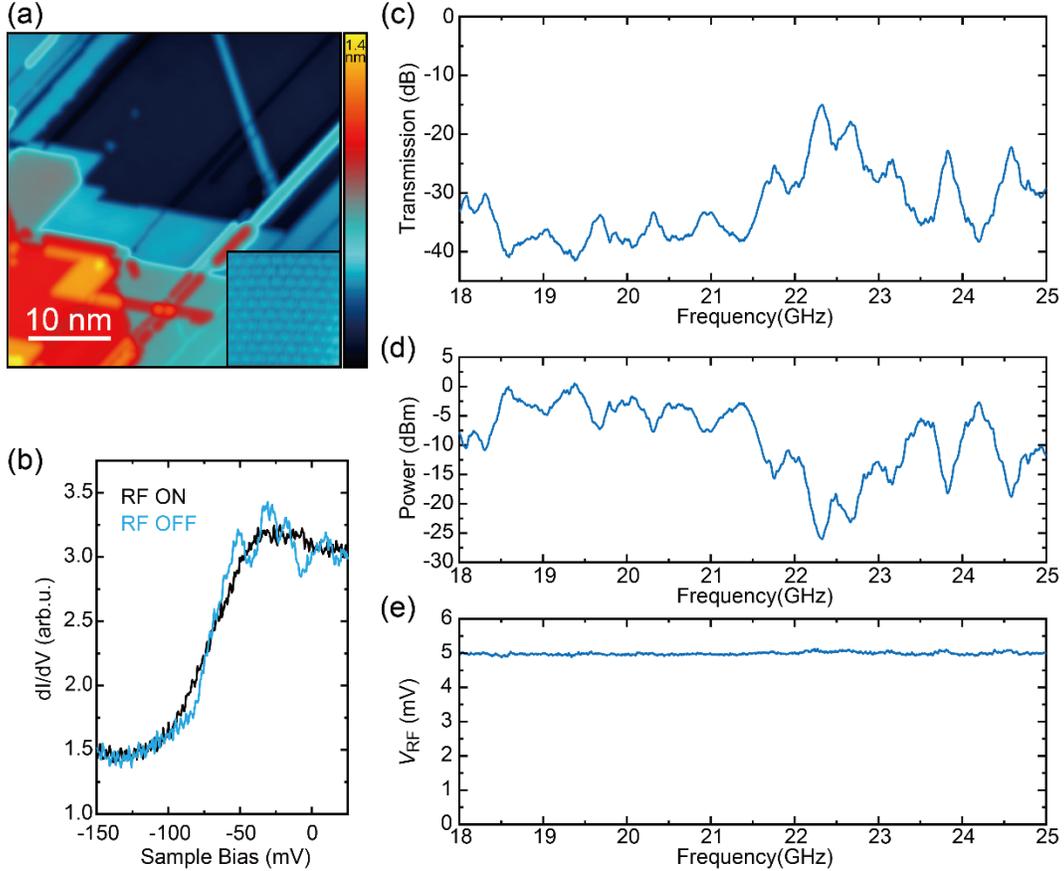

**Figure S3.** Compensation of RF transmission at the tunneling junction. (a) Typical large-scale STM image of Ag(111) surface (scanning condition: $V_{DC} = 0.9$ V and $I = 18$ pA). The inset shows an image at an atomic resolution (scanning condition: $V_{DC} = 70$ mV and $I = 500$ pA). (b) $dI/dV$ spectra of step-like Ag(111) surface state with (black curve) and without (blue curve) RF signal ($P_{RF} = -5$ dBm, $f = 19$ GHz). The applied continuous RF signal broadened the step-like surface state of Ag(111) owing to the application of the $V_{RF}$ at the tunnel junction; the $V_{RF}$ amplitude was estimated to be 25 mV. (c) RF transmission of frequency range 18–25 GHz at a constant RF source power equal to $P_{RF} = 5$ dBm. (d) The RF source power used to compensate the frequency-dependent RF transmission and create constant $V_{RF}$ at tunnel junction. (e) The constant $V_{RF}$ amplitude of 5 mV was achieved by utilizing a compensated source power.

## S4. Statistical analysis of the number of isolated TbPc2 molecules on the surface

As shown in Figure 1(b) in the main text, there are several isolated single TbPc$_2$ molecules over the Cu(100) surface, while a few isolated single molecules exist on the NaCl surface where most molecules either aggregate at the edge of the NaCl island or make clusters. The same tendency was observed in other areas of the same sample as shown in Figure S4. We estimated the densities of the isolated molecule on the Cu and NaCl surfaces to be 17 molecules per 1618 nm$^2$ ~ $1.1 \times 10^{-3}$ nm$^{-2}$, and 2 molecules per 482 nm$^2$ ~ $4.1 \times 10^{-3}$ nm$^{-2}$, respectively. We performed the same analysis in ten different areas, including those shown in Figures 1b and S4. The averaged density ratio of the isolated molecules found on the Cu and NaCl surfaces was equal to 3:1.

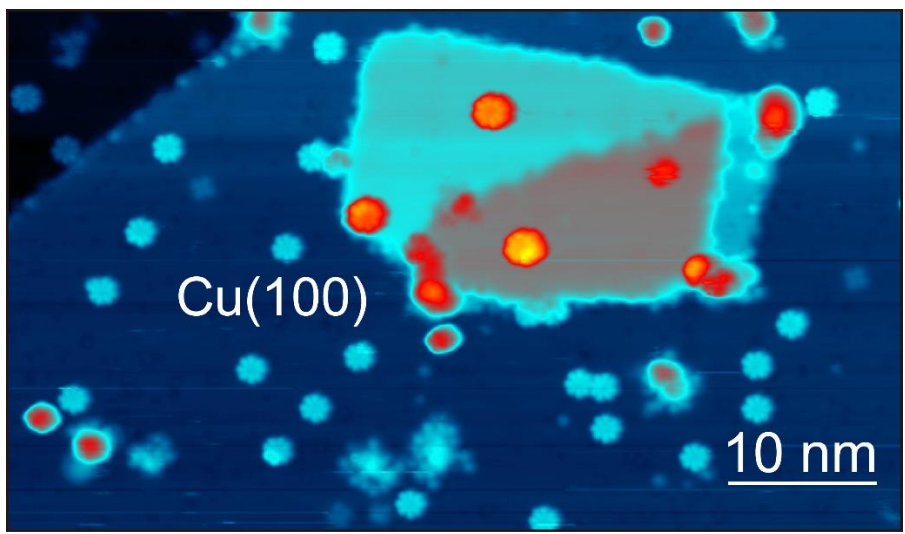

**Figure S4.** STM image of TbPc2 molecules adsorbed on bare Cu(100) and 2 ML NaCl/Cu(100) surfaces. Scanning conditions: $V_{DC}$ = 1.4 V and $I$ = 10 pA.